\documentclass[preprint]{aastex}

\shortauthors{A.K.Inoue, H.Hirashita and H.Kamaya}
\shorttitle{Star Formation Efficiency}

\begin{document}

\title{Star Formation Efficiency in the Central 1 kpc Region of Early-Type
Spiral Galaxies}

\author{Akio K. Inoue, Hiroyuki Hirashita\altaffilmark{1} and Hideyuki
Kamaya}

\affil{Department of Astronomy, Faculty of Science, Kyoto University,
Sakyo-ku, Kyoto 606-8502, JAPAN}

\email{AKI: inoue@kusastro.kyoto-u.ac.jp}

\altaffiltext{1}{Research Fellow of Japan Society for the Promotion of
Science.}

\begin{abstract}

It has been reported recently that there are some early-type spiral
 (Sa--Sab) galaxies having evident star-forming regions which
 concentrate in their own central 1-kpc.
In such central region, is the mechanism of the star formation distinct
 from that in disks of spiral galaxies? 
To reveal this, we estimate the star formation efficiency (SFE) in this
 central 1-kpc star-forming region of some early-type spiral galaxies,
 taking account of the condition for this 1-kpc region to be
 self-gravitating.
Using two indicators of present star formation rate (H$\alpha$ and
 infrared luminosity), we estimate the SFE to be a few percents.
This is equivalent to the observational SFE in the disks of late-type
 spiral (Sb--) galaxies.
This coincidence may support the universality of the mean SFE of spiral
 galaxies reported in the recent studies.
That is, we find no evidence of distinct mechanism of the star formation 
 in the central 1-kpc region of early-type galaxies.
Also, we examine the structure of the central star-forming region, and
 discuss a method for estimating the mass of star-forming regions.

\end{abstract}

%\keywords 

%\newpage

\section{INTRODUCTION}

When we try to reveal the galaxy evolution along with the theory of star
and star-cluster formation, it is very important to examine the 
star formation activity (SFA) of galaxies.
The SFA has been frequently examined by using the star formation rate
(SFR), which is the mass turned into stars per unit time.
Also, the star formation efficiency (SFE) is an important indicator 
of the SFA.
Here, we define the SFE as being the ratio of the stellar mass in a
star-forming region to the gas mass of a parent cloud.
This SFE relates more directly than the SFR to the mechanism of turning
the interstellar matter into stars.
Thus, we focus our attention on the SFE in galaxies.

In this paper, we are interested in early-type spiral galaxies
(Sa--Sab), since they are reported to indicate an interesting property
of star formation.
The SFA in such galaxies is frequently considered to be 
lower than that in late-type spiral (Sb--) galaxies (e.g., \citealt{kea98}).
However, there are some early-type spiral galaxies showing evident
star-forming  activity (e.g., \citealt{kee83, kee85, ket89, you96,
tom96, dev97, ham99}).
Interestingly, star-forming regions of such early-type spiral galaxies
concentrate in their own central 1-kpc region \citep{usu98}.
According to Inoue, Hirashita \& Kamaya (2000, hereafter IHK00), the SFR 
of this central 1-kpc star-forming region is about 2 $M_\sun$ yr$^{-1}$,
that is, the SFR in early-type spirals is comparable to that in late-disks.
In order to obtain a more precise picture of the SFA in those central
regions, we should examine their SFE.
Thus, the main aim of this paper is to estimate the SFE of the central
1-kpc star-forming region in early-type spiral galaxies.

In order to estimate the SFE of a star-forming region, we must examine
both of the stellar mass and the gas mass within this region.
We will estimate the stellar mass from the H$\alpha$ luminosity by using
the formula of the SFR in \cite{kea98} and an appropriate time-scale
for the current star formation.
We also estimate it from the {\it IRAS}-far-infrared (FIR)
luminosity via the formula of IHK00 which is reviewed in Appendix A.
By the way, 
when we estimate the mass of molecular gas of the sample galaxies, there
is no available complete data set for them, unfortunately.
Therefore, we estimate the gas mass of the central star-forming
regions, taking account of the condition for these regions to be
self-gravitating.
This is because if the system were not to be self-gravitating,
the cite for the star formation could be disturbed by external
gravity, so that the star formation itself would be terminated.
Also, we discuss the structure of the central star-forming region, and
present a method for estimating the gas mass of star-forming regions.

We compare the estimated SFE with typical observational SFEs 
in the disk of late-type spiral galaxies.
As a result, we will show that the SFE in the central part of early-type
spiral galaxies is equivalent to that of the late-type disks.
This result may support the universality of the mean SFE of spiral galaxies
reported by \cite{you96} and \cite{row99}.
We might need to expect an universal mechanism for star formation
in spiral galaxies.

This paper contains the following sections. First, we examine the
structure of the central star-forming region in early-type spiral
galaxies in \S 2.
Then, we estimate the SFE of this region in \S 3.
Next, we compare our SFE with other observational ones and discuss the
critical surface density for self-gravitational contraction 
to star formation in \S 4.
Conclusions are summarized in the final section.

\section{STRUCTURE OF CENTRAL STAR-FORMING REGION}

In this paper, we will discuss the physical condition of the
central 1-kpc region of the early-type spiral galaxies. Hence,
first of all, we review the observational picture of those central
regions.  Moreover, in this section,
we discuss and examine the structure of the star-forming region within the
central 1-kpc. A simple structural model for this region is proposed.

\subsection{Observational properties}

\cite{usu98} observed H$\alpha$ and $R$-band photometric images (the
field of view was $10' \times 6'$) of nearby 15 early-type spiral (Sa--Sab)
galaxies which show relatively high star formation activity (SFA)
determined from the ratio of {\it IRAS}-FIR luminosity to $B$-band
luminosity.
Some properties of their sample galaxies are tabulated in
Table~\ref{tbl-1}.
They commented that their observed equivalent widths of the
H$\alpha$ emission line agree with those measured by
\citet{ken83} and \citet{rom90}.

When we consider that H$\alpha$ emission originates from
H {\footnotesize II} regions where massive stars have formed,
the half-light radius of H$\alpha$ emission represents a characteristic
radius of the spatial distribution of star-forming regions.
{}From Table~\ref{tbl-1}, we find that the half-light radii of
H$\alpha$ are about 1 kpc (col.\ [7]).
In addition, we find from column 8 that the radii of H$\alpha$ are much
smaller than the optical radii which are one-half of the optical
diameter, $D_{25}$, in \cite{dev91} (RC3).
Thus, we expect that the star-forming regions in such galaxies
distribute in their central $\sim$ 1 kpc region and
concentrate relatively to their optical stellar light.

Next, let us look at kinematical properties of the central 1-kpc region
in spiral galaxies.
Recently, \cite{sof99} investigated rotation curves of 50 nearby
late-type spiral (Sb--Sc) galaxies, using position-velocity diagrams
composed of both observations of CO and H {\footnotesize I} lines.
This method is suited to determine the central rotation curves.
According to them, typical spiral galaxies show almost flat rotation
curves from their central a few hundreds pc to their outer edge.
That is, the rotation of a typical spiral galaxy is differential in its
central 1-kpc region.
\cite{rub85} examined the rotation curves of early-type spirals via the
observation of H$\alpha$ line.
Then they concluded that the rotation curves of early-type spirals are
similar to those of late-type spirals.
Thus, we regard the rotation of the central 1-kpc star-forming region in 
early-type spiral galaxies as being also differential.

Let us turn to the subject of the gas content in the central region.
According to \cite{you91}, who presented a detailed review of radial gas
distribution, the molecular gas concentrates in the central several kpc
region of spiral galaxies, while the H {\footnotesize I} gas
distributions are rather flat throughout their disk.
Moreover, the molecular fraction of hydrogen is almost unity in the central
a few kpc of nearby spiral galaxies \citep{sof95, hon95}.
Also, a dense molecular disk has been observed in the center of the
Galaxy \citep{san84}.
Furthermore, a diffuse coronal gas component has been observed at
X-ray in many spiral galaxies (e.g., \citealt{rea97}).
Thus, we can expect that the gas content of the central 1 kpc region is a
mixture of the two components; one of the components is a dense molecular
gas and the other is a diffuse hot gas.

\subsection
{Characteristic Length Scale of a Differentially Rotating Medium}

To make the following discussion clear, as only a first step, 
we assume that the components of interstellar medium (ISM)
in the star-forming regions are
well mixed enough for us to regard these regions as uniform.
Hereafter, in this paper, `uniform medium' or `homogeneous medium'
is sometimes used to denote the well-mixed ISM.
Also, we assume that these regions rotate differentially and are
established with a global pressure balance of the ISM \citep{mye78}.
If the mixture of some ISM components is well relaxed to equilibrium
via pressure, 
there is a characteristic sound speed which is defined
as being the root of the  ratio of the mean pressure to
the mean density of the system 
(e.g., \citealt{kam98}).
This definition is possible when the thermal process is very active
between the two components (e.g., \citealt{kam99}).
Moreover, we consider that the radius of this region is larger than the
characteristic scale of its self-gravity, the Jeans length,
since stars and clouds do form in the regions.
Then, we present a formula of the Jeans length for our purpose.
 
For an infinite and homogeneous gaseous system rotating
differentially, \cite{bel58} and \cite{gen75} have examined the
self-gravitational instability of the direction perpendicular to the
rotational axis in a cylindrically symmetric geometry.
According to them, the dispersion relation of a small perturbation in
such a system is expressed by the following,
\begin{equation}
 \omega^2 = \frac{4 \pi^2 {c_s}^2}{\lambda^2} - 4\pi G \rho + \kappa^2\,,
 \label{disp}
\end{equation} 
where $\rho$ and $c_s$ denote the mean density and the characteristic 
sound speed of this system, respectively, and $\omega$ and $\lambda$
are the
frequency and the wavelength of the perturbations, respectively.
Moreover, $\kappa$ represents an epicyclic frequency of the rotation of
this system and it is defined by
\begin{equation}
 \kappa^2 \equiv 4\Omega^2 + R\frac{d\Omega^2}{dR}\,,
\end{equation}
where $\Omega$ is an angular speed of the system and $R$ is a radius
from the rotational axis.
If the system has a constant rotational speed, $v_{\rm rot}$ (i.e. a
differential rotation), then, $\kappa^2=2{v_{\rm rot}}^2/R^2$.

When $\omega^2<0$, the system becomes unstable owing to its self-gravity.
The minimum wavelength that satisfies this condition is called 
the Jeans length.
Therefore, we obtain the Jeans length, $\lambda_J$,
\begin{equation}
 \lambda_J = \frac{2\pi c_s}{\sqrt{4\pi G \rho - \kappa^2}}\,.
 \label{jlen1}
\end{equation}

Now, we consider an uniform medium obeying a global pressure balance.
The sound speed, $c_s$, is expressed by ${c_s}^2
\approx p/\rho$ and the global pressure balance is represented by
$p/k_{\rm B}\approx a$ (a constant), where $p$ is the pressure of this 
medium, and $k_{\rm B}$
denotes the Boltzmann constant.
Thus, equation~(\ref{jlen1}) is reduced to 
\begin{equation}
 {\lambda_J}^2 = \left( \frac{\pi a k_{\rm B}}{G} \right)
                 \left[ \rho 
                 \left( \rho - \frac{{v_{\rm rot}}^2}{2\pi GR^2} \right)
                 \right]^{-1}\,.
 \label{jlen2}
\end{equation}
Although we find that $a$ is $10^{3-4}$ K cm$^{-3}$ from \cite{mye78},
we adopt $a$ as $10^4$ K cm$^{-3}$ in this paper, because the central
region seems to have a bit higher pressure than the other regions and
hot gas component has several $10^{-3}$ H atoms cm$^{-3}$ as its number
density and several $10^6$ K as its temperature (e.g., \citealt{rea97}).

\subsection{Required Mean Density for 1-kpc Scale Contraction}

The SFA in the early-type spiral galaxies may be
explained by the gravitational contraction of the central 1-kpc region.
The radius of a star-forming region, $r_{\rm SF}$, should be
more than $\lambda_{\rm J}$ to contract by self-gravity, that is
$r_{\rm SF}>\lambda_{\rm J}$.
{}From equation~(\ref{jlen2}), we determine 
the following condition for its contraction:
\begin{equation}
 \rho^2 - A \rho - B > 0\,,
 \label{dens}
\end{equation}
where
\begin{equation}
 A = \frac{{v_{\rm rot}}^2}{2\pi GR^2} \quad
 {~~\rm and ~~~}
 B = \frac{\pi a k_{\rm B}}{G {r_{\rm SF}}^2}\,.
\end{equation}
Since $\rho>0$, we obtain the solution of inequality~(\ref{dens}): 
\begin{equation}
 \rho > \frac{A}{2} + \frac{A}{2}\sqrt{1+\frac{4B}{A^2}}\,.
\end{equation}
When $r_{\rm SF}\sim$ 1 kpc \citep{usu98} and $v_{\rm rot}=200$ km
s$^{-1}$ (a typical value for normal size galaxies in Sofue et
al.~1999), we obtain the following relation,
\begin{equation}
 \frac{4B}{A^2} \simeq 2.7\times10^{-3}\left(\frac{R}{\rm kpc}\right)^4
                \ll 1 \,,
\end{equation}
as far as $R \la 3$ kpc, which is satisfied for the typical
central star-forming regions in early-type spiral galaxies.
Thus, the above condition is reduced to $\rho >A$.

Consequently, we determine a condition of the density for gravitational
contraction on 1-kpc scale as a function of the radius, and define 
a critical density, $\rho_{\rm crit}$,
\begin{equation}
 \rho(R) > \frac{{v_{\rm rot}}^2}{2\pi GR^2} \equiv \rho_{\rm crit}\,.
 \label{density}
\end{equation}
Since the mean density of the central star-forming region in early-type
spiral galaxies should be more than this critical density and the size
of the region is estimated to be about 1 kpc, 
a minimum mean density required for the contraction is
\begin{equation}
 \langle\rho\rangle_{\rm min} = \rho_{\rm crit}(1~{\rm kpc}) 
          \sim 1.0\times10^{-22}\, [{\rm g\,cm^{-3}}]\,, 
\end{equation}
or the corresponding number density is
\begin{equation}
 \langle n \rangle_{\rm min} \sim 60\,[{\rm H\,atoms\,cm^{-3}}]\,, 
\end{equation}
when we adopt 200 km s$^{-1}$ as the rotational speed.

\subsection{Structural Model of Central Star-Forming Region}

Someone might think that the mean density estimated in the
previous subsection is too large
since the main component of the ISM in the central regions is just the
hot gas bound in a deep gravitational potential well of these regions.
Fortunately, however, as commented in \S2.1, 
there are molecular clouds in the central regions,
although the spatial distribution of the molecular clouds
are not precisely determined. 
Hence, we examine if this minimum density of equation (10) or (11)
can be reproduced easily by the two components of the ISM.
For example, when the number density of the component of molecular gas
is $10^3$ H atoms cm$^{-3}$ and that of hot gas is several times
$10^{-3}$ H atoms cm$^{-3}$, we reproduce the minimum density 
if we adopt only 0.06 as the volume
filling factor of molecular gas. This small filling factor of 0.06 
never contradicts to the observations of molecular clouds
in the central regions.

Here, we examine if the two components are well mixed 
via a dynamical process before the star formation onsets. 
Fortunately, this is possible, because while the sound crossing
time-scale over 1 kpc is about $10^7$ years in hot gas, the
gravitational time-scale in the minimum density is at most about $10^7$
years.
We want to note here that
the overall contraction with the Jean-length is very slow  
(we should never forget this point), since
the growth-rate of the mode with Jeans length is zero theoretically
(in a realistic condition, of course, the zero growth-rate breaks
due to the variety of disturbances).
Thus, the two components can be well mixed when the system begins
to contract. 
The effect of the differential motion can also mix the two components,
since the rotation time-scale is also about 10$^7$ years.
To conclude, our estimate in equations (10) and (11) is very reasonable 
since it is possible for the two components to be well mixed.
The above estimation of time-scales confirms our another assumption that
the global pressure balance establishes in this region before
the star formation onsets, since the molecular clouds are
well mixed into the hot gas.

{}From the above discussions, we regard that the 1 kpc star-forming
regions in the center of early-type spiral galaxies are the regions composed
of the mixture of the molecular and coronal gas components.
The treatment as a self-gravitational contraction of the mixture
in 1-kpc scale becomes suitable
when their high mean 
density with the volume filling factor of the cold gas of
$\sim 0.06$ is realized. The minimum density for it can be reached
and thus the contraction of the system may occur, 
when the mixture suffers the radiative cooling. 
We comment again on the requirement for the system to be self-gravitating.
If the self-gravity of the system were to be less important,
the cite for the star formation could be disturbed very well 
by the external gravitational force.
As a result, the star formation would be terminated there.

\section{STAR FORMATION EFFICIENCY}

We estimate the star formation efficiency (SFE) in the central 1-kpc
region of early-type spiral galaxies showing evident star-forming activity.
We define the SFE of a star-forming region as the ratio of the stellar
mass newly formed in the region ($M_*$) to its total gas mass ($M_{\rm
gas}$).
That is,
\begin{equation}
 {\rm SFE} \equiv \frac{M_*}{M_{\rm gas}}\,.
 \label{sfe}
\end{equation}
In the following subsections, we estimate both $M_{\rm gas}$ (\S 3.1) and
$M_*$ (\S 3.2), so that we determine the SFE (\S 3.3).

\subsection{Estimation of Gas Mass}
We first consider the gas mass of the star-forming region.
Now, we want to examine the SFA of the 1-kpc regions.
Since the thickness of a typical galactic disk is about 100 pc, we 
cannot consider that such a star-forming region is a sphere.
Therefore, the global structure of the star-forming regions are assumed
to be a disk.

In order to estimate the gas mass in such disk-like regions, we must
consider their thickness.
We mean here the term of the thickness as that of the direction parallel
to the rotational axis.
When the thickness is denoted by $h_{\rm SF}$, the gas mass of a
star-forming region is given by
\begin{equation}
 M_{\rm gas} = \int_0^{r_{\rm SF}} 2\pi r\, h_{\rm SF}\rho\, dr\,,
\end{equation}
where $r$ means the radius of the star-forming region extended
to $r_{\rm SF}$ and $\rho$ is its mean density.

Since the star-forming regions are self-gravitating, their
thickness is more than the Jeans length.
In the direction of the thickness, the term of the epicyclic frequency
in equation~(\ref{jlen1}) does not exist.
Thus, the thickness of the star-forming regions, $h_{\rm SF}$, should
satisfy the following condition:
\begin{equation}
 h_{\rm SF} > \sqrt{\frac{\pi {c_s}^2}{G\rho}}
      = \frac{1}{\rho}\sqrt{\frac{\pi a k_{\rm B}}{G}}\,,
 \label{thick}
\end{equation}
where we use again ${c_s}^2 \approx ak_{\rm B}/\rho$, which is
obtained under
the assumption that the well-mixed ISM obeys a global pressure
balance (\S 2.2).

{}From inequality~(\ref{thick}), we find that the minimum of $h_{\rm
SF}\rho$ depends on only $a$, that is,
\begin{equation}
 h_{\rm SF}\rho > \sqrt{\frac{\pi a k_{\rm B}}{G}}\,.
 \label{sfcrit}
\end{equation}
If we assume that the global pressure ($ak_{\rm B}$) does not vary
significantly in the central star-forming regions, we obtain the 
condition for the gas
mass in such star-forming regions as the following,
\begin{equation}
 M_{\rm gas} > \pi {r_{\rm SF}}^2 h_{\rm SF} \rho
             = \pi {r_{\rm SF}}^2 \sqrt{\frac{\pi a k_{\rm B}}{G}}\,,
 \label{gasmass1}
\end{equation}
where ${r_{\rm SF}}$ is the radius or the size of a star-forming region.
When we adopt $10^4$ K cm$^{-3}$ as $a$, we obtain the following condition;
\begin{equation}
 \frac{M_{\rm gas}}{M_\sun} 
 > 1.2 \times 10^8 \left( \frac{r_{\rm SF}}{\rm kpc} \right) ^2 \,.
 \label{gasmass2}
\end{equation}
Therefore, we can estimate the minimum gas mass of the central
star-forming regions through inequality~(\ref{gasmass2}) under the
assumption that a global pressure balance is established
in this region.

{}From the observation of \cite{usu98}, we adopt 1 kpc as the radius of
the central star-forming region in early-type spiral galaxies.
Therefore, we estimate the characteristic gas mass of such regions to
be more than $1.2\times10^8M_\sun$.
Since it is reported that the mass of molecular gas within the central
1.5 kpc of the Galaxy is about $5\times10^8M_\sun$ \citep{san84}, the
estimated gas mass may be reasonable.
Also, since the central molecular mass for four sample galaxies has been
reported, we discuss them individually in \S 3.4.

\subsection{Estimation of Stellar Mass}

\subsubsection{From the H$\alpha$ Luminosity}
The H$\alpha$ luminosity is the most familiar tracer of young massive
stars.
Thus, we estimate the stellar mass newly born from the H$\alpha$
luminosity.
According to \cite{kea98}, the formula to estimate the SFR from the
H$\alpha$ luminosity, $L_{\rm H\alpha}$, is
\begin{equation}
 \frac{\rm SFR}{M_\sun\,{\rm yr^{-1}}} 
 = 3.05 \times 10^{-8} \frac{L_{\rm H\alpha}}{L_\sun}\,.
\end{equation}
When a typical time-scale of the star formation, $t_{\rm SF}$, is given
and the SFR does not vary significantly during this time-scale, the
newly born stellar mass, $M_*$, is estimated to be 
\begin{equation}
 M_* = {\rm SFR} \times t_{\rm SF}\,.
\end{equation}

In this paper, we consider the time-scale, $t_{\rm SF}$, to be a
luminosity-weighted average lifetime of the main-sequence OB stars
which are representative of the young population, since we
are interested in the current star formation rate to obtain
the present SFE.
Thus, we regard the time-scale of the star formation as a typical
main-sequence lifetime of OB stars.
IHK00 have estimated the lifetime to be $3.3\times10^6$ yr, adopting
their mass-luminosity relation, main-sequence lifetime as a function 
of stellar mass, and Salpeter's IMF.
We adopt their value here (i.e., $t_{\rm SF}=3.3\times10^6$ yr).
Moreover, we assume the SFR to be nearly constant during this 
timescale because it is much shorter than the timescale of the
galaxy evolution (typically $10^{9}$ yr).
Therefore, the stellar mass expected from the H$\alpha$ luminosity is 
\begin{equation}
 \frac{M_*}{M_\sun} = 0.10 \frac{L_{\rm H\alpha}}{L_\sun}\,.
 \label{starmass}
\end{equation}

The observed luminosities of H$\alpha$ + [N {\footnotesize II}] line of sample galaxies in
\cite{usu98} are tabulated in Table~\ref{tbl-1}.
Their averaged luminosity is
$(3.4\pm 3)\times10^7L_\sun$\footnote{Throughout this paper, the numbers
after $\pm$ mean the typical dispersions of the sample data}.
To obtain a reasonable luminosity due to only starformation, 
we correct it for the contamination of [N {\footnotesize II}] line 
and the internal
extinction (mean A$_{\rm H\alpha} \sim 0.8$ mag from the private
communication with Usui).
The corrected H$\alpha$ luminosity is $(4.8\pm 4)\times10^7L_\sun$.
Consequently, we estimate the stellar mass to be 
$(4.8\pm 4)\times10^6M_\sun$.

\subsubsection{From the Infrared Luminosity}
We can also estimate the stellar mass newly born in galaxies from their
infrared (IR) luminosity.
There are many formulae to estimate it from IR luminosity as, for example,
\cite{ken98}.
Unfortunately, his formula can be applied to only starburst galaxies
because he adopted an assumption that IR luminosity of galaxies is almost
equal to their bolometric one.
We call this assumption the ``starburst approximation.''
Studies of this kind often adopted this
``starburst approximation'' (otherwise an {\it empirical} relation
between the luminosities of IR and another wavelength is adopted;
for example,
\citealt{buat96} used the luminosity ratio of IR to ultraviolet).
However, this approximation cannot be applied to the galaxies in the
sample of \cite{usu98}.
Thus, a new {\it theoretical} formula to estimate the 
stellar mass from IR luminosity must be applied to this sample.
IHK00 have constructed a new algorithm to do so.
Their algorithm is reviewed briefly in Appendix A.
The derived formula is
\begin{equation}
 \frac{M_*}{M_\sun}=\frac{1.5 \times 10^{-3}(1-\eta)}
                         {0.4-0.2f+0.6\epsilon}\,
                    \frac{L_{\rm FIR}(40-120\mu{\rm m})}
                         {L_\sun}\,,
\end{equation}
where $f$ is a fraction of ionizing photons absorbed by hydrogen atoms
in H {\footnotesize II} regions, $\epsilon$ denotes an efficiency of
dust absorption for nonionizing photons, and $\eta$ represents a cirrus
fraction.
Furthermore, $L_{\rm IR}^{\rm obs}(8-1000\mu{\rm m})$ in equation~(\ref{a3})
is converted to {\it IRAS}-FIR luminosity, $L_{\rm FIR}(40-120\mu{\rm
m})$, by the factor of 1.4, assuming that the radiation from dust is the 
modified black body radiation of 30 K ($I_\nu \propto \nu B_\nu$, where
$I_\nu$, $\nu$, and $B_\nu$ are intensity, frequency, and the Plank
function, respectively).

The available data set in \cite{usu98} has {\it IRAS}-FIR (40--120
$\mu$m) luminosity of 15 early-type spiral galaxies, and it is tabulated in
Table~\ref{tbl-1}.
The FIR luminosity of this sample is about
$(8.0\pm 6)\times10^9L_\sun$.
We estimate a characteristic stellar mass newly formed in the central
1-kpc region of this sample by using this averaged FIR
luminosity.
Of course, this FIR luminosity originates from the whole galaxy.
For the sample of early-type spiral galaxies, however, the SFA in their disks
seems to be very low.
Thus, after we subtract the ``cirrus'' component, which originates from
dust distributed diffusely in the interstellar space and unrelated to
star-forming regions, from the observed FIR luminosity, we assume that
the whole of residual FIR luminosity originates in the central 1 kpc
star-forming region only.

Here, we choose three parameters as the following (see Appendix B):
$f=0.26$, $\epsilon=0.6$, and $\eta=0.5$.
Since the FIR luminosity of these sample galaxies is
$(8.0\pm 6)\times10^9L_\sun$, then, $M_*=(8.5\pm 6)\times10^6M_\sun$.
This is consistent with the stellar mass estimated from the H$\alpha$
luminosity.
Thus, we have confirmed the applicability of the formula of IHK00.

\subsection{Estimation of SFE}
We now estimate a characteristic SFE of the central 1-kpc star forming
region in the sample galaxies of \cite{usu98}.
We have already estimated that the gas mass is heavier
than $1.2\times 10^8 M_{\odot}$ in \S 3.1 and 
that the stellar mass from the H$\alpha$ luminosity is 
$(4.8\pm 4)\times10^6M_\sun$ in the previous subsection.
Therefore, we obtain
\begin{equation}
 {\rm SFE_{\rm H\alpha}} < 0.04\pm0.03\,.
\end{equation}

Moreover, we can estimate the SFE from the FIR luminosity.
As seen in \S3.2.2, the stellar mass determined by the FIR luminosity
is $(8.5\pm 6)\times10^6M_\sun$.
Thus, we obtain 
\begin{equation}
 {\rm SFE_{\rm FIR}} < 0.07\pm0.05\,.
\end{equation}
Here, we find a good agreement between the both estimations.
However, we should note that the way of estimating from the H$\alpha$
luminosity may be more reliable than that from the FIR.
This is because the estimation from the FIR luminosity suffers more
severe uncertainties, for example, the ``cirrus'' fraction and
the spatial distribution of the FIR luminosity.

Anyway, we conclude that the characteristic SFE expected in the central
1-kpc star forming regions in the early-type spiral galaxies is at most
a few percents, which is comparable to SFEs in galactic disks as
reviewed in \S 4.1.

\subsection{Individual SFE of Four Sample Galaxies}
For four of our sample galaxies, the central molecular masses have
been reported.
In Table~2, the observed molecular masses are tabulated.
Thus, we can determine directly their central SFE.

In column~(3) of Table~2, we show the radii of the area
integrated CO intensity to estimate the molecular mass.
These CO radii, $r_{\rm CO}$, are more extended than those of their 
star-forming region, $r_{e,\rm H\alpha}$ in Table~1.
Hence, we determine the gas mass of the star forming regions by 
\begin{equation}
 M_{\rm gas} = \left(\frac{r_{e,\rm H\alpha}}{r_{\rm CO}}\right)^2
               M_{\rm H_2}\,.
 \label{obsgas}
\end{equation}
Consequently, we obtain the following equation from equations
(\ref{sfe}), (\ref{starmass}), and (\ref{obsgas});
\begin{equation}
 {\rm SFE} = 0.10\left(\frac{r_{\rm CO}}{r_{e,\rm H\alpha}}\right)^2
             \frac{L_{\rm H\alpha}/L_\sun}{M_{\rm H_2}/M_\sun}\,.
 \label{SFE}
\end{equation}

As shown in column~(5) of Table~2, the determined SFEs are about one
percent.
We find a good agreement between the both SFEs estimated in \S 3.3 and
here.
Therefore, our estimation in the previous subsection is quite reasonable.

\section{DISCUSSIONS}

In this section, we discuss some implications derived from our results
in the previous sections. 
Especially, we are interested in
whether our SFE is larger or smaller than
other SFEs in various conditions of star formation.
If our SFE is similar to the mean value, the star formation
mechanism, which is typically a mode in molecular clouds, may be
universal everywhere, except for starburst, strongly interacting,
and so on.
Thus, we shall examine the physical reason why our SFE becomes a few
percents.
For this purpose, we also examine the criterion
of the contraction of the parent clouds for star formation.

\subsection{Comparison with Other Observational SFEs}
We compare the SFE estimated in \S 3.3 with other SFEs determined
observationally.
\cite{mye86} examined 54 molecular clouds associated with H
{\footnotesize II} regions in the Galactic disk.
These molecular clouds locates in $-1^\circ \leq b \leq 1^\circ$ and
$12^\circ \leq l \leq 60^\circ$.
They estimated the molecular mass of each cloud from
observations of CO line emission, and also estimated the stellar mass in
H {\footnotesize II} regions accompanying each cloud from H 110$\alpha$
radio (6 cm) emission line.
Their result indicates that the averaged SFE is about 0.02.
\cite{wil97} also argued that the Galactic averaged SFE is about 0.05.
Moreover, the SFEs of two giant H {\footnotesize II} regions (NGC 604 and
NGC 595) in the disk of M33 are reported to be 0.02--0.05 by 
\cite{wil95}.
Therefore, we find that the SFE expected in the central 1-kpc
star-forming region of early-type spiral galaxies is very similar to that in
the disk of late-type spiral galaxies.

By the way, we can also define a kind of the SFE of galaxies from the
ratio of their luminosity of newborn stars to their molecular mass.
When we consider that H$\alpha$ luminosity, $L_{\rm H\alpha}$, traces
the mass of recently formed stars, $L_{\rm H\alpha}/M_{\rm H_2}$ is
regarded as an
indicator of the SFE, where $M_{\rm H_2}$ is the molecular mass.
\cite{you96} and \cite{row99} observed H$\alpha$ and $R$-band 
luminosities of spiral galaxies in the sample of
the Five College Radio Astronomy
Observatory (FCRAO) {\it Extragalactic CO Survey} \citep{you95}, and
examined morphological variations of the SFE for their sample by using
this ratio.
Then, they concluded that
there is little variation in the mean SFE among the morphologies
of the spiral galaxies and it is about 4\%.
As a result, the coincidence between our theoretical SFE in the center of
early-type spiral galaxies and observational ones in the disk of late-type spiral
galaxies may support the universality of the mean SFE against the
morphological variation of spiral galaxies.
This may indicate that the same process such as contraction of
molecular clouds by their self-gravity dominates the star
formation in all spiral galaxies.

\subsection{Criterion of Self-Gravitational Contraction}
Here, we first discuss inequality~(\ref{sfcrit}).
The left-hand side of it is the thickness, $h_{\rm SF}$, multiplied by
the density, $\rho$, of star-forming regions.
This represents a surface density of these regions.
Thus, inequality~(\ref{sfcrit}) suggests a criterion of the surface
density for the self-gravitational contraction on the assumption that
the medium in these regions is well mixed and obeys the law of global
pressure balance.

We rewrite this criterion of the surface density, defining 
a critical surface density, $\Sigma_{\rm crit}$;
\begin{equation}
 h_{\rm SF}\rho > \sqrt{\frac{\pi a k_{\rm B}}{G}} \equiv
\Sigma_{\rm crit}\,.
\end{equation}
When we adopt $a=10^4$ K cm$^{-3}$ as the law of
the global pressure balance (\S 2.2), 
we obtain the following value,
\begin{equation}
 \Sigma_{\rm crit}\sim 40~[M_\sun\,{\rm pc}^{-2}]\,. 
 \label{sigma1}
\end{equation}
It is the most important point that the critical surface density depends 
on only the global pressure represented by $a$.
Once $a$ is given, the gas mass of a star-forming region is reduced to a
function of its radius through inequality~(\ref{gasmass2}).

By the way, the global pressure balance is a balance among various
components of the ISM in a steady state.
Thus, the ISM evolving dynamically does not always obey this balance.
For example, the H {\footnotesize II} regions, 
located inside the molecular clouds with active star formation, 
are not in an equilibrium state but evolve dynamically.
Therefore, it is very interesting to examine more precisely
how the global pressure balance among the ISM components
establishes. 
Then, we shall compare our critical surface density with an alternative one.

In the 1-kpc star-forming regions, we also consider that the thickness of
these regions is more than the Jeans length and the density is also more
than the critical density in equation~(\ref{density}).
Thus, we find another critical surface density, $\Sigma_{\rm crit}'$;
\begin{equation}
 h_{\rm SF}\rho > \sqrt{\frac{\pi {\sigma}^2 \rho_{\rm crit}}{3G}}
                = \frac{\sigma}{\sqrt{6}\, G}\,\frac{v_{\rm rot}}{R}
                \equiv \Sigma_{\rm crit}'\,,
\end{equation}
where we eliminate $\rho_{\rm crit}$, using equation~(\ref{density}),
and a characteristic sound speed is rewritten by $(\sigma ^2/ 3)^{1/2}$
in terms of a velocity dispersion, $\sigma$, of the ISM mixture.
Therefore, we obtain the following equation,
\begin{equation}
 \frac{\Sigma_{\rm crit}'}{M_\sun\,{\rm pc}^{-2}} 
 = \frac{\sigma}{\sqrt{6}\, G}\,\frac{v_{\rm rot}}{R}
 \simeq 0.6\left(\frac{\sigma}{\rm 6.0~km\,s^{-1}}\right)
           \left(\frac{v_{\rm rot}}{\rm km\,s^{-1}}\right)
           \left(\frac{\rm kpc}{R}\right)\,,
 \label{sigma2}
\end{equation}
where we have adopted 
$\sigma\sim6$ km s$^{\rm -1}$, according to \cite{ken89}.
This is consistent with the critical surface density given in
equation (6) of
\cite{ken89}, who derived it from the criterion of instability for a
thin isothermal disk of \cite{too64}.

We thus find that the estimated $\Sigma_{\rm crit}$ in
equation~(\ref{sigma1}) is nearly equal to $\Sigma_{\rm crit}'$ in
equation~(\ref{sigma2}) when we adopt 200 km s$^{-1}$ as $v_{\rm rot}$.
Now, we try to find the reason for this coincidence between
equations~(\ref{sigma1}) and (\ref{sigma2}).
To determine equation~(\ref{sigma1}), we adopt $p\sim 10^{-12}$ erg
cm$^{-3}$, then for $\rho_{\rm crit} \sim 10^{-22}$ g cm$^{-3}$,
we find a characteristic sound speed is an order of km s$^{-1}$.
By the way, it seems that $\sigma$ is also an order of km s$^{-1}$.
This can mean that $\sigma$ is decided by the way
how ISM is well mixed and is in global pressure balance. 
Thus, our critical surface density of equation~(\ref{sigma1})
coincides with equation~(\ref{sigma2}) 
and with \citet{ken89}'s result from the criterion of \cite{too64}, 
since there exists the law of the global pressure balance 
in the well-mixed ISM.
Anyway, our conclusion is based on the two physical assumptions;
(a) there is a global law of the pressure balance among the ISM;
(b) the ISM components are well mixed among each other.
The first assumption is established via Figure 1 of \cite{mye78},
and the second one is good as far as 
the effective crossing time is not longer than the contraction
time-scale of the starforing disk as discussed in \S 2.4.

\section{CONCLUSIONS}

We examine the SFE of the central star-forming region in early-type spiral
galaxies, and reach the following conclusions:

[1] We estimate the SFE in the central 1-kpc star-forming region of some
early-type spiral galaxies, taking account of the condition for this 1-kpc
region to be self-gravitating.
The expected SFE is at most a few percents.

[2] The SFE of four sample galaxies (NGC1022, NGC2782, NGC3504, and
NGC7625), whose central molecular mass is measured observationally,
is estimated to be about one percent.

[3] The estimated SFE is equivalent to the observational ones in the disk
of late-type spiral galaxies.
This coincidence may support the universality of the mean SFE against the
morphological variation of the spiral galaxies.

[4] Assuming a global pressure balance of the ISM, the critical surface
density for the self-gravitational contraction depends on 
the adopted pressure balance.
As a result, the mass of star-forming regions is always estimated via
only their size once the pressure balance is given.

[5] We suggest that the
central 1 kpc star-forming region of early-type spiral galaxies is composed
of the mixture of the molecular and coronal gas components. 
The mixture is possible to contract in 1 kpc scale via their
self-gravity if the mean gas density of the mixture is 
more than a minimum value (e.g., 60 H atoms cm$^{-3}$ in this paper).

[6] The formula of IHK00 is reasonable since
SFE estimated from far-infrared luminosity is nearly equivalent
to that determined from H$\alpha$ luminosity.

\acknowledgments

We would like to thank Prof.\ M.\ Sait\={o} for
continuous encouragement.
One of us (H.H.) acknowledges the Research Fellowship of
the Japan Society for the Promotion of Science for Young
Scientists.

\appendix

\section{Newly Formed Stellar Mass Expected from Infrared Luminosity}
We present a formula to estimate the stellar mass newly formed in
galaxies from their IR luminosity of dust emission.
More detailed information can be found in \cite{ihk00} (hereafter
IHK00) who construct a new algorithm to estimate the SFR from IR
luminosity.
The result of IHK00 is an extension of a theory of IR radiation from a
dusty H {\footnotesize II} region in \cite{pet72}.

\cite{pet72} derived a formula of IR luminosity of an H {\footnotesize
II} region, adopting the Case B approximation, which is an assumption of
a large optical depth for every Lyman-emission-line photon \citep{ost89},
and assuming that all Lyman $\alpha$ photons are absorbed by dust grains
in the H {\footnotesize II} region.
Subsequently, IHK00 has developed this result of \cite{pet72}, adopting
the Salpeter's IMF, $\psi(m)\propto m^{-2.35}$ for
$0.1M_\sun<m<100M_\sun$, a stellar mass-luminosity relation fitted to Table
3.13 of \cite{bin98} and a function of the luminosity fraction of
ionizing photons from young massive stars fitted to Table 2 in
\cite{pan73}.
They also assume that the luminosity of star-forming regions 
is dominated by the bolometric luminosities of OB stars on the main-sequence.
Moreover, observed IR luminosity of galaxies consists of two components: 
[1] the component originating from dust nearby star-forming regions and
[2] the component originating from dust unrelated to such regions and
distributed diffusely in the interstellar space, called the cirrus
component (e.g., \citealt{hel86,lon87}).
Thus, IR luminosity of star-forming regions is the luminosity subtracted 
this cirrus component from observed IR luminosity.
As a result, they obtain the luminosity of star-forming regions,
\begin{equation}
 L_{\rm SF}=\frac{1-\eta}{0.4-0.2f+0.6\epsilon}\,L_{\rm IR}^{\rm obs}\,,
 \label{a1}
\end{equation}
where $f$ is a fraction of ionizing photons absorbed by hydrogen atoms
in H {\footnotesize II} regions, $\epsilon$ denotes an efficiency of
dust absorption for nonionizing photons and $\eta$ represents a cirrus
fraction.

On the other hand, the stellar mass newly formed in star-forming regions 
expected from their luminosity is expressed by
\begin{equation}
 \frac{M_*}{M_\sun}=\frac{\displaystyle \int_{0.1M_\sun}^{100M_\sun}m\psi(m)dm}
                   {\displaystyle \int_{3M_\sun}^{100M_\sun}l(m)\psi(m)dm}\,
                    \frac{L_{\rm SF}}{L_\sun}
                   =1.1 \times 10^{-3}\, \frac{L_{\rm SF}}{L_\sun}\,,
 \label{a2}
\end{equation}
where $m$ denotes a stellar mass in solar unit, $l(m)$ means the adopted 
mass-luminosity relation as a function of a stellar mass in solar unit,
and the mass range of OB stars is adopted 3--100$M_\sun$.
{}From equation~(\ref{a1}) and (\ref{a2}), IHK00 has obtained the stellar
mass expected from observed IR luminosity as the following formula,
\begin{equation}
 \frac{M_*}{M_\sun}=\frac{1.1 \times 10^{-3}(1-\eta)}
                         {0.4-0.2f+0.6\epsilon}\,
                    \frac{L_{\rm IR}^{\rm obs}(8-1000\mu{\rm m})}
                         {L_\sun}\,.
 \label{a3}
\end{equation}
Here, the spectral range of observed IR luminosity is considered to be
$8-1000\mu{\rm m}$ as the almost whole range of the thermal dust
radiation.

Then, IHK00 estimate a typical time-scale of the star formation which is
luminosity-weighted average lifetime of the main-sequence OB stars, so
that they derive the formula estimating the SFR as the following,
\begin{equation}
 \frac{{\rm SFR}}{M_\sun\, {\rm yr^{-1}}}= \frac{M_*}{t_{\rm SF}}=
  \frac{3.3\times 10^{-10}(1-\eta)}
       {0.4-0.2f+0.6\,\epsilon}\,
  \frac{L{\rm ^{obs}_{IR}}(8{-}1000\,\mu{\rm m})}{L_\sun}\,,      
 \label{a4}
\end{equation}
where $t_{\rm SF}=3.3\times10^6$ yr as the time-scale of the star
formation.

\section{Determination of Three Parameters}
Here, we determine values of three parameters in equation~(\ref{a3}).
For absorbing fraction by neutral hydrogen, $f$, we adopt 0.26 derived by
Petrosian et al. (1972) from an optical depth of dust in Orion nebula.
The efficiency of dust absorption for nonionizing photons,
$\epsilon$, is estimated to be 0.6 from the averaged 1000--4000\AA\
extinction curve of the Galaxy (Savage \& Mathis 1979) and the
average visual extinction of Usui's sample ($A_V = 1$ mag from private
communication with Usui).
Here, we should note that when our formula is applied to
the individual molecular clouds, these parameters may depend on the geometry 
of clouds.
Moreover, we choose 0.5 for the cirrus fraction, $\eta$, of Usui's sample,
according to Lonsdale-Persson \& Helou (1987), who presented a model of
the cirrus fraction by using the ratio of IRAS 60$\mu$m and 100$\mu$m
fluxes.
Also, we can estimate this fraction by applying a proper radiative
transfer model for reproducing multi-wavelength data of galaxies (e.g.,
Silva et al.~1998, Efstathiou et al.~2000).
However, the fraction of the cirrus component remains still uncertain.

\clearpage

\begin{deluxetable}{lccccccc}
\footnotesize
\tablecaption{Some Properties of Sample Galaxies of \cite{usu98} \label{tbl-1}}
\tablewidth{0pt}
\tablehead{
\colhead{Galaxy} & \colhead{$B^0_T$} & \colhead{$\log{L_{\rm FIR}}$} &
 \colhead{$f_{60}/f_{100}$} & \colhead{$\log{L_{\rm FIR}/L_B}$} &
 \colhead{$\log{L_{\rm H\alpha + [N {\footnotesize II}]}}$} &
 \colhead{$r_{e,\rm H\alpha}$}  & \colhead{$r_{e,\rm H\alpha}/r_{25}$} \\
\colhead{} & \colhead{} & \colhead{($L_\sun$)} & \colhead{} & \colhead{} &
 \colhead{($L_\sun$)} & \colhead{(kpc)}  & \colhead{} \\ 
\colhead{(1)} & \colhead{(2)} & \colhead{(3)} & \colhead{(4)} &
 \colhead{(5)}  & \colhead{(6)} & \colhead{(7)} & \colhead{(8)} \\
}
\startdata
NGC681  & 12.50 & 9.49 & 0.36 & $-0.23$ & 7.16 & 2.5 & 0.28 \\
NGC1022 & 11.94 & 10.08 & 0.75 & $~~0.29$ & 7.12 & 0.4 & 0.06 \\
NGC2782 & 12.01 & 10.26 & 0.59 & $~~0.03$ & 7.84 & 1.0 & 0.06 \\
NGC2993 & 12.65 & 10.22 & 0.63 & $~~0.36$ & 7.91 & 0.7 & 0.12 \\
NGC3442 & 13.64 & 9.16 & 0.46 & $-0.07$ & 7.10 & 0.7 & 0.33 \\
NGC3504 & 11.51 & 10.17 & 0.65 & $~~0.20$ & 7.66 & 0.6 & 0.07 \\
NGC3611 & 12.31 & 9.48 & 0.61 & $-0.15$ & 7.09 & 0.6 & 0.09 \\
NGC3729 & 11.91 & 9.08 & 0.36 & $-0.45$ & 7.15 & 1.5 & 0.26 \\
NGC4045 & 12.52 & 9.88 & 0.52 & $~~0.12$ & 7.13 & 2.2 & 0.21 \\
NGC4369 & 12.27 & 9.32 & 0.54 & $-0.05$ & 7.02 & 0.5 & 0.11 \\
NGC4384 & 13.38 & 9.66 & 0.44 & $-0.04$ & 7.33 & 1.2 & 0.18 \\
NGC5534 & 12.68 & 9.91 & 0.69 & $-0.06$ & 7.73 & 1.8 & 0.26 \\
NGC5691 & 12.52 & 9.55 & 0.57 & $-0.19$ & 7.60 & 1.2 & 0.18 \\
NGC5915 & 11.99 & 10.19 & 0.67 & $~~0.05$ & 7.93 & 1.5 & 0.18 \\
NGC7625 & 12.67 & 9.95 & 0.52 & $~~0.29$ & 7.47 & 0.9 & 0.17 \\
\hline
average & 12.4$\pm1.1$ & 9.9$^{+0.2}_{-0.6}$ & 0.6$\pm0.2$ & $~~0.01\pm0.2$ &
 7.5$^{+0.3}_{-0.9}$ & 1.2$\pm0.6$ & 0.2$\pm0.2$ \\
\enddata

\tablecomments{Col.(1): Galaxy name. Col.(2): $B^0_T$ from the
 RC3. Col.(3): $\log{L_{\rm FIR}}$ is calculated using $f_{60}$ and
 $f_{100}$ from {\it IRAS} data in solar units. Col.(4): Ratio of
 $f_{60}$ and $f_{100}$. Col.(5): $\log{L_{\rm FIR}/L_B}$ is calculated
 using cols.(2) and (3). Col.(6): Observed H$\alpha$ + [N {\footnotesize
 II}] luminosity by \cite{usu98}. Col.(7): Half-light radius of
 H$\alpha$ + [N {\footnotesize II}] emission in kiloparsecs. Col.(8):
 Ratio of the half-light radius of H$\alpha$ + [N {\footnotesize II}]
 emission to the optical radius, $r_{25}$, which is one-half of the
 optical diameter in the RC3. We assume $H_0=75$ km s$^{-1}$
 Mpc$^{-1}$. The standard deviation is also shown in the average of each
 column.}

\end{deluxetable}

\clearpage

\begin{deluxetable}{lccccl}
\footnotesize
\tablecaption{Individual Properties of Four Sample Galaxies with CO
 Observation}
\tablewidth{0pt}
\tablehead{\colhead{Galaxy} & \colhead{$M_{\rm H_2}$} 
 & \colhead{$r_{\rm CO}$} & \colhead{$L_{\rm H\alpha}$} & \colhead{SFE}
 & \colhead{CO Reference}\\
\colhead{} & \colhead{($10^9M_\sun$)} & \colhead{(kpc)} &
 \colhead{($10^7L_\sun$)} & \colhead{(\%)} & \colhead{}\\ 
\colhead{(1)} & \colhead{(2)} & \colhead{(3)} & \colhead{(4)} &
 \colhead{(5)} & \colhead{(6)}\\}

\startdata
NGC1022 & 1.5 & 1.65 & 1.8 & 2.0 & \cite{gar91}\\
NGC2782 & 2.5 & 1.3~ & 9.5 & 0.6 & \cite{jog99}\\
NGC3504 & 1.8 & 0.8~ & 6.3 & 0.6 & \cite{ken93}\\
NGC7625 & 2.4 & 1.75 & 4.1 & 0.6 & \cite{li93}\\
\enddata

\tablecomments{Col.(1): Galaxy name. Col.(2): Molecular mass estimated
 from CO intensity via the CO/H$_2$ conversion, $2.8\times10^{20}$ H$_2$ 
 cm$^{-2}$/K km s$^{-1}$. Col.(3): Radius of the integrated area to
 estimate molecular mass. Col.(4): H$\alpha$ luminosity corrected for
 the contamination of [N {\footnotesize II}] line and
 extinction. Col.(5): SFE determined by equation~(\ref{SFE}).}

\end{deluxetable}


\begin{thebibliography}{}
\bibitem[Bel \& Schatzman(1958)]{bel58}
        Bel, N., \& Schatzman, E.  1958, RvMP, 30, 1015

\bibitem[Binney \& Merrifield(1998)]{bin98}
        Binney, J., \& Merrifield, M.  1998, Galactic Astronomy (New
        Jersey: Princeton University Press) chap.3

\bibitem[Buat \& Xu(1996)]{buat96} Buat, V., \& Xu, C. 1996, \aap,
        306, 61

\bibitem[de Vaucouleurs et al.(1991)]{dev91}
        de Vaucouleurs, G., de Vaucouleurs, A., Corwin, H. G., Buta, R. J.,
        Patural, G., Fouqe\'{e}, P.  1991, Third Reference Catalogue of
        Bright Galaxies (New York: Springer) (RC3)

\bibitem[Devereux \& Hameed(1997)]{dev97}
        Devereux, N. A., \& Hameed, S.  1997, \aj, 113, 599

\bibitem[Efstathiou et al.(2000)]{efs00}
        Efstathiou, A., Rowan-Robinson, M., \& Siebenmorgen, R.  2000,
        \mnras, 313, 734

\bibitem[Garcia-Barreto et al.(1991)]{gar91}
        Garcia-Barreto, J. A., Downes, D., Combes, F., Gerin, M.,
        Carrasco, L., \& Cruz-Gonzales, I.  1991, \aap, 252, 19

\bibitem[Genkin \& Safronov(1975)]{gen75}
        Genkin, I. L., \& Safronov, V. S.  1975, SvA, 19, 189

\bibitem[Hameed \& Devereux(1999)]{ham99}
        Hameed, S., \& Devereux, N. 1999, \aj, 118, 730

\bibitem[Helou(1986)]{hel86}
        Helou, G. 1986, \apjl, 311, L33

\bibitem[Honma et al.(1995)]{hon95}
        Honma, M., Sofue, Y., \& Arimoto, N.  1995, \aap, 304, 1

\bibitem[Inoue, Hirashita \& Kamaya(2000)]{ihk00}
        Inoue, A. K, Hirashita, H., \& Kamaya, H.  2000, \pasj, 52, 539
	(astro-ph/0003318) (IHK00) 

\bibitem[Jogee et al.(1999)]{jog99}
        Jogee, S., Kenney, J. D. P., \& Smith, B. J.  1999, \apj, 526, 665

\bibitem[Kamaya(1999)]{kam99}
        Kamaya, H. 1999, PASJ, 51, 617 

\bibitem[Kamaya \& Shchekinov(1998)]{kam98}
        Kamaya, H. \& Shchekinov, Yu. A., 1998, PASJ, 50, 621 

\bibitem[Keel(1983)]{kee83}
        Keel, W. C.  1983, \apj, 268, 632

\bibitem[Keel et al.(1985)]{kee85}
        Keel, W. C., Kennicutt, R. C. Jr., Hummel, E., \& van der
        Hulst, J. M.  1985, \aj, 90, 708

\bibitem[Kenney et al.(1993)]{ken93}
        Kenney, J. D. P., Carlstrom, J. E., \& Young, J. S. 1993, \apj,
        418, 687

\bibitem[Kennicutt(1983)]{ken83}
        Kennicutt, R. C. Jr.\ 1983, \apj, 272, 54

\bibitem[Kennicutt(1989)]{ken89}
        Kennicutt, R. C. Jr.\ 1989, \apj, 344, 685

\bibitem[Kennicutt(1998a)]{kea98}
        Kennicutt, R. C. Jr.\ 1998a, \araa, 36, 189

\bibitem[Kennicutt(1998b)]{ken98}
        Kennicutt, R. C. Jr.\ 1998b, \apj, 498, 541

\bibitem[Kennicutt et al.(1989)]{ket89}
        Kennicutt, R. C. Jr.\, Edgar, B. K., \& Hodge, P. W.  1989, \apj,
        337, 761

\bibitem[Li et al.(1993)]{li93}
        Li, J. G., Seaquist, E. R., Wrobel, J. M., Wang, Z. \& Sage,
        L. J.  1993, \apj, 413, 150

\bibitem[Lonsdale-Persson \& Helou(1987)]{lon87}
        Lonsdale-Persson, C. J., \& Helou, G.  1987, \apj, 314, 513
 
\bibitem[Myers(1978)]{mye78}
        Myers, P. C.  1978, \apj, 225, 380

\bibitem[Myers et al.(1986)]{mye86}
        Myers, P. C., Dame, T. M., Thaddeus, P., Cohen, R. S.,
        Silverberg R. F., Dwek, E., \& Hauser, M. G.  1986, \apj, 301,
        398

\bibitem[Osterbrock(1989)]{ost89}
        Osterbrock, D. E.  1989, Astrophysics of Gaseous Nebulae \&
        Active Galactic Nuclei (Mill Valley: University Science Books)
        chap.4

\bibitem[Panagia(1973)]{pan73}
        Panagia, N. 1973, \aj, 78, 929

\bibitem[Petrosian et al.(1972)]{pet72}
        Petrosian, V., Silk, J., \& Field, G. B.  1972, \apjl, 177, L69  

\bibitem[Read et al.(1997)]{rea97}
        Read, A. M., Ponman, T. J., \& Strickland, D. K.  1997, \mnras,
        286, 626

\bibitem[Romanishin(1990)]{rom90}
        Romanishin, W. 1990, \aj, 100, 373

\bibitem[Rubin et al.(1985)]{rub85}
        Rubin, V. C., Burstein, D., Ford, W. K., \& Thonnard, N  1985,
        \apj, 289, 81

\bibitem[Rownd \& Young(1999)]{row99}
        Rownd, B. K., \& Young, J. S.  1999, \aj, 118, 670

\bibitem[Sanders et al.(1984)]{san84}
        Sanders, D. B., Solomon, P. M., \& Scoville, N. Z.  1984, \apj,
        276, 182

\bibitem[Silva et al.(1998)]{sil98}
        Silva, L., Granato, G. L., Bressan, A., \& Danese, L.  1998, \apj,
        509, 103

\bibitem[Sofue et al.(1995)]{sof95}
        Sofue, Y., Honma, M., \& Arimoto, N.  1995, \aap, 296, 33

\bibitem[Sofue et al.(1999)]{sof99}
        Sofue, Y., Tutui, Y., Honma, M., Tomita, A., Takamiya, T., Hodo, 
        J., \& Takeda, Y.  1999, \apj, 523, 136

\bibitem[Tomita et al.(1996)]{tom96}
        Tomita, A., Tomita, Y, \& Sait\={o}, M.  1996, \pasj, 48, 285

\bibitem[Toomre(1964)]{too64}
        Toomre, A.  1964, \apj, 139, 1217

\bibitem[Usui et al.(1998)]{usu98}
        Usui, T., Sait\={o}, M., \& Tomita, A.  1998, \aj, 116, 2166

\bibitem[Williams \& McKee(1997)]{wil97}
        Williams, J. P., \& McKee, C. F.  1997, \apj, 476, 166

\bibitem[Wilson \& Matthews(1995)]{wil95}
        Wilson, A. D., \& Matthews, B. C.  1995, \apj, 455, 125

\bibitem[Young et al.(1996)]{you96}
        Young, J. S., Allen, L, Kenney, J. D. P., Lesser, A., \& Rownd, 
        B.  1996, \aj, 112, 1903

\bibitem[Young \& Scoville(1991)]{you91}
        Young, J. S., \& Scoville, N. Z.  1991, \araa, 29, 581

\bibitem[Young et al.(1995)]{you95}
        Young, J. S., Xie, S., Tacconi, L., Knezek, P., Viscuso, P.,
        Tacconi-Garman, L., Scoville, N., Schneider, S., Schloerb,
        F. P., Lord, S., Lesser, A., Kenney, J., Huang, Y. -L.,
        Devereux, N., Claussen, M., Case, J., Carpenter, J., Berry, M.,
        \& Allen, L.  1995, \apjs, 98, 219

\end{thebibliography}
\end{document}